\documentclass[a4paper,12pt]{article}

\begin{document}
\title{Statistics, ethics, and probiotica\footnote{{\it Statistica Neerlandica} (2009) Vol. 63, nr. 1, pp. 1--12}}
\author{Richard D.~Gill
\\
Mathematical Institute
\\
Leiden University
\\
{\small http://www.math.leidenuniv.nl/$\sim$gill; gill@math.leidenuniv.nl}
}
\date{4 September 2008}
\maketitle
\begin{abstract}

\noindent Ethical issues involved in the design of the ``PROPATRIA'' probiotica trial are discussed. This randomized clinical trial appeared to be well conducted according to accepted good practices. The finding that the treatment was actually rather harmful, and that despite this, and despite a built-in interim analysis, the trial was not stopped earlier, led to strong criticism in the media. 

I argue that ``accepted good practices'' need to be reconsidered in the light of this experience. Firstly, a much stronger distinction needs to be recognized between the immediate interests of the patients being treated in the trial, and the interests of future patients of future doctors elsewhere. Secondly, it is in the interests of future patients that well conducted clinical trials are accepted by society. Since it is unavoidable that an occasional trial will result in an unpredicted severely negative outcome, ethical screening committees must ensure that those performing a trial can never be accused of putting the interest of ``science'' above the interest of their own patients when such ``accidents'' happen.

There are two consequences of this. Firstly, the design of a trial should also explicitly minimize the number of patients who are treated by the researchers with a potentially seriously harmful medicine. Secondly, the disadvantages of triple-blinding far outweigh the advantages. Though it might at best only have saved a few lives if the PROPATRIA trial been re-designed with these issues in mind, I argue that the scientific value of the trial would not have been significantly reduced; the damage to medical research, and hence to future patients, would have been substantially less. 

Closer inspection of the data from the PROPATRIA trial brings a new and quite unexpected failing to light. The decision for early stopping the trial was accidentally based on the one-sided test looking in the wrong direction, partly through inadequacy of the output of the statistical package SPSS, partly through lack of statistical expertise on the part of the users.  If the envisaged one-sided stopping rule had been used correctly, the trial would in fact have been terminated at the time of the interim analysis ``for futility'': it was at this moment highly unlikely that a significant end-result in favour of probiotica was going to be attained. The decision to continue the trial was a result of looking at the test statistic ``in the wrong direction''. In effect, the trial was continued because there was still a good chance to show that probiotica is actually very harmful.

I recommend that data monitoring committees should always be advised by a professional statistician and that this person is not blinded to the treatment allocation.

\end{abstract}

\section{The Design}

The recently concluded PROPATRIA study of the use of probiotics in the treatment of acute pancreatitis (Besselink et al., 2004, 2008) received much media attention in the Netherlands when it was revealed at the close of the study that the treatment appeared to have a negative effect. It did not, as expected, decrease the chance of infectious complications, which remained at around 1 in 3. More seriously, among those that did develop infectious complications, the frequency of a fatal outcome was roughly doubled (roughly, from 1 in 4 to 1 in 2). 

For ethical reasons the multi-centre, doubled-blinded, randomized trial had been monitored over the the three years of its duration, and an interim analysis performed. The interim analysis was further semi-blinded in the sense that the monitoring committee did not know which group was the treatment group, which group was the control. Though, at the final conclusion of the trial,  there had been a total of 33 deaths in the total group of 296 patients considered in the final analysis, only two of these deaths had been considered by doctors in the participating hospitals as so unexpected and serious at the time, that they were reported individually to the monitoring committee. The treatment group was revealed to the monitoring committee for those two cases only.  In both cases the patient was in the treatment group; and in both cases the death was due to the same rather unusual complication which -- only retrospectively -- turned out to be the cause of a large number of the probiotica deaths.

The research team conducting the trial certainly strongly believed at the outset that probiotica likely would have a good effect on their patients. They had theoretical arguments supporting that opinion, and also empirical evidence pointing in the good direction (but there have also been negative reports). Moreover, the treatment consists of a cocktail of bacteria which, among healthy people, are normal and necessary residents of our digestive tract, and which are present in popular and commercially available food additives. New strains had been developed which were expected to stimulate immune response, and to compete with the ``bad'' bacteria associated with the infectious complications which are the usual cause of mortality in pancreatitis. The researchers considered that there were good reasons to carry out a large scale trial to settle the issue, and did not expect any negative consequences of the treatment.

In retrospect, the triple blinding of the study is obviously an embarrasment to the already seriously disappointed and concerned researchers. This is especially the case since the trial was originally planned to include only 200 patients, not 300\footnote{Actually: 298 patients entered the trial, but in retrospect, it appeared 2 had been wrongly diagnosed and are removed from the analysis. The final report says that the monitoring committee advised increasing to 296. I wonder if the monitors really wrote this number, or if it was a slip of the pen of the final reporter, two years later. Whether this is exactly the number which came out of a calculation or not, I replace it here by a round number}. At the interim analysis of the first 100, the overall rate of infectious complication was lower than expected. The monitoring committee made the recomendation to increase the total intake to 300 patients ``in order to preserve statistical power''.  This is an intervention on behalf of science, not on behalf of the patients in the trial, as I will further argue below.  The formal interim analysis was postponed to the new, expected, half-way mark, but actually only took place when the treatment of a total of 184 patients had been completed. The mortality of the two groups seemed different but the difference did not quite reach significance at the 10\% level. The monitoring committee associated this with a slight difference in the initial health state of the two groups; and the monitoring committee did not know which group was which. Presumably, if anything they could imagine that the trial was working out -- to the advantage of the treatment -- as the researchers had initially expected. An early stopping rule based on the ``primary endpoint'' of the trial, infectious complications, seemed to indicate that the trial should continue.

The statistical design was based on mostly standard recipes distilled from a standard work -- at least, locally, in the Netherlands --- Schouten (1999). That author highly recommends the interim analysis advised by Snapinn (1992). A small poll of professors of medical statistics in the Netherlands revealed that despite Schouten's endorsement, this method is more or less forgotten. Internationally the methods of choice seem to be those of Pocock, and of Fleming and O'Brien. However \emph{The Lancet} follows Sir Richard Peto's much more radical advice only to act on interim analysis significance tests if they reach the 1 in 1000 significance level: in other words, almost never. There is a good philosophy behind this recommendation, but it is based on assumptions, and the question is always, whether those assumptions ought to have been made in this case, even if they might be usually uncontroversial.

Triple blinding is very controversial, and is much discussed in the literature.  The FDA extensively discuss the pros and cons in their guidelines for clinical trials in drug testing, and advises against it. A Dutch epidemiologist Vandenbroucke (1999) endorses it strongly in a three page article in the \textit{Nederlands Tijdschrift voor de  Geneeskunde}. He gives good reasons for it, but there are also reasons against it, and the question is, what is most relevant for this specific case.  Normal practice in the US and the UK is that even if the data monitoring committee is blinded, the statistician who is advising them is not blinded. He or she is able to intervene if the committee seems to be making a decision which they would regret if the identity of the two groups were exchanged. Normal practice is that the data monitoring committee explicitly plays through both scenarios in their deliberations: group A is treatment, B is control; and vice versa. If their decision under both scenarios is the same the data does not need to be de-blinded.

In the paper Besselink et al.~(2004) describing the design of the trial in \textit{BMC Surgery}, the researchers write

\begin{quote}
This study is conducted in accordance with the principles 
of the Declaration of Helsinki and `good clinical practice'.

For ethical reasons it is desirable to end a therapeutic 
experiment once a statistical significant difference in treatment 
results has been reached. This study uses the stopping-rules 
according to Snapinn (1992). An interim-analysis 
will be performed after the data of the first 100 patients 
(50\% fraction) is obtained. According to Snappin, the 
trial will be ended at this interim-analysis at $p < 0.0081$. 
The study will also be ended in case of adverse events 
without possibility of positive outcome, $p > 0.382$. The 
monitoring committee will discuss the results of the 
interim-analysis and advice the steering committee. The 
steering committee decides on the continuation of the 
trial.
\end{quote}

Now, there are certainly ethical reasons to conclude a trial as soon as possible. Schouten (1999) gives a list of four ethical concerns (and more can be added):
\begin{itemize}
\item[1.] Is it ethical to give a placebo treatment to a seriously ill person?
\item[2.] Is it ethical to give an untested new medicine, which might have serious side effects, to a seriously ill person?
\item[3.] Is it ethical to decide treatment for a seriously ill patient, with the interests of science in mind, on the basis of tossing a coin?
\item[4.] Is it ethical still not to know, 10 or 20 years from now, what is the best treatment for a seriously ill patient?
\end{itemize}

The point I want to make is that \textit{different ethical concerns are in conflict with one another. Any proposed ``solution'' implictly weights the different concerns in a particular way}. 

Let me first sketch the global design of the trial. It was set up to have power 0.80 when testing the null hypothesis of no treatment effect, two sidedly, and with significance level 0.05, against the alternative that the treatment roughly halved the probability of the ``primary endpoint'', infectious complications. This means that the primary constraint on the researchers is to guard against the publication of ``false positives''. They mustn't run a higher risk than 1 in 20 of this. And journals which publish the results of medical research, insist on the same ``filtering'' of signal from noise.  

Since the power against the actually expected effect is 0.80, they are prepared to run a risk of 1 in 5 that the treatment cannot be proved effective, even if it is exactly as effective as they believe.  It is perhaps hard to believe that one invests so much research effort with a 1 in 5 chance that it is all wasted, but this seems to be standard practice. On the other hand, one could say that taking such a low sample size is actually protecting patients, in the case that the new treatment turns out to be bad for them.

Whose ethical concerns are addressed by these choices? Primarily, the concerns of science and of future patients of future doctors. We don't want to tell the world that probiotica is fantastic, when actually it does nothing. (And \textit{The Lancet} won't let us do this either). There is an unpleasant side effect here: if probiotica actually doubles the risk of infectious outcomes, instead of halving it, we also run a risk of 1 in 5 of not noticing that, and just getting a non-significant result. At least, a false-positive result does not get published in that case either. Would \textit{The Lancet} have published a non-significant result anyway? It is surely bad to tell the world that probiotica makes no difference, when it actually doubles your risk.

Now let us overlay these, already ethical choices, with the choice implied by Snapinn's method. 

The reason Schouten is so enthusiastic about Snapinn is that it does not alter the evaluation of the final results when the trial is not stopped half-way. Thus the nominal size $\alpha=0.05$ is maintained, even though there is some chance that the sample size is half what it apparently should have been.

There are no free lunches. This means that \textit{either} the power is reduced, \textit{or} the chance to stop early hardly exists.

Well, we have to be a bit more careful, since the chance of early stopping depends on ``the truth''. Snapinn is conservative, and does not want to lose much power. There is some chance of early stopping if the null hypothesis is true, but almost no chance of early stopping if the alternative hypothesis is true. Thus the power is only slightly reduced. Schouten has adapted Snapinn's design, and accepts a much larger loss of power in order to increase the chances of early stopping, under both hypotheses.

Whose ethical concerns are addressed by use of this early stopping rule? Clearly it is good to stop early if probiotica doesn't do anything at all (though whether \textit{The Lancet} would still publish if the trial is aborted halfway because nothing is expected to come of it, is an interesting question). What about the issues of interest to the patients actually entering the trial? If a new treatment has unpleasant side effects but otherwise makes no difference, then the patients who would normally enter the trial later, would appreciate not receiving it. It's a bit strange to call the ``secondary outcome'' death, merely an unpleasant side effect. 

And how big is the chance of stopping early, if the treatment has no effect at all? The answer is easy to read off, from the actual implementation of the Snapinn rule:  the researchers planned to stop the trial half-way, if the $p$-value at that time was larger than 0.382. This means that there was a 62\% chance that the trial would be stopped half-way if actually the treatment did not reduce infectious complications (as indeed seems to be the case). If the treatment has a negative effect, the probability of stopping early ``for futility'' rapidly gets larger. So at this point the trial does have a built-in safety measure.

Whose ethical concerns are addressed by adopting Vandenbroucke's advice to triple blind? This makes it even less likely for a monitoring committee to ``break the rules'' by stopping the trial early when it is going in an unexpected, negative direction. The point is that future patients of future doctors have an interest in trials being completed to full term, otherwise results are biased. A trial which is stopped early because of an apparent negative effect is probably stopped when the observed (negative) effect is worse than the actual one. 

In conclusion: ethical safeguards were built into the statistical design of the trial, but they are a variety safety measures for different, conflicting, ethical issues. Some safety measures actually increase the ethical dangers in the situation which by the admission of the researchers themselves, most likely obtained for the probiotica trial -- namely testing a treatment which turned out to be harmful for their patients. For this eventuality, just one statistical safety measure was built in -- the possibility of ``stopping for futility'' in the Snapinn plan.

\section{The Results}

In their 2008 publication in \textit{The Lancet} at the close of the trial, the researchers write

\begin{quote}
We calculated that 200 patients with predicted severe acute 
pancreatitis would be required to detect a 20\% reduction in 
the absolute risk of the occurrence of infectious 
complications (from 50\% to 30\% of patients during 
admission and 90-day follow-up) for the study to attain an 
80\% statistical power, at a two-sided $\alpha$ of 0.05. This sample 
size calculation took into account the fact that up to 40\% of 
patients with predicted severe pancreatitis are ultimately 
diagnosed with mild pancreatitis (i.e., no local or systemic 
complications) and thus do not progress to severe or 
necrotising pancreatitis. After the first 100 patients were 
randomised and had completed follow-up, the number of 
infectious complications was calculated in the total group.

The rate of infectious complications was lower than 
expected (28\%), so the monitoring committee advised 
increasing the total sample size from 200 to 296 patients to 
maintain statistical power. After 184 patients had been 
randomised and had completed follow-up, a blinded 
interim analysis was done for the primary endpoint and 
mortality. Although a non-signiÞcant difference in 
mortality was observed ($p=0.10$), the monitoring committee 
concluded that this had been caused by skewed 
randomisation because more patients in the group with 
higher mortality required admission to intensive care 
within 72 hours after admission ($p=0.15$), whereas the overall 
mortality was well within the expected range (11\%). 
According to the predefined stopping rule the monitoring 
committee recommended that the study should be 
completed.

During the study, two serious adverse events 
were reported; both patients died. The monitoring 
committee convened on both occasions: in one patient, a 
ruptured caecum with ischaemia was found during 
emergency laparotomy and the second patient had 
small-bowel ischaemia diagnosed at emergency laparotomy. 
In both cases, the randomisation code was broken (both 
patients had received probiotics). This information was 
revealed only to members of the monitoring and steering 
committees. A review of published work did not reveal any 
evidence of a relation between bowel ischaemia and the 
use of probiotics. The monitoring committee subsequently 
advised that the study continue. The institutional review 
board was informed on both occasions.

\end{quote}

I want to draw attention to two things: firstly, the advice of the monitoring committee to increase the sample size in order to maintain statistical power. Actually, this possibility was envisaged in the original protocol of the trial, for the following reason. Patients had to be entered into the trial on the basis of \emph{predicted} severe acute pancreatitis. It takes another week or more before a more certain diagnosis can be made, but it was important to start the treatment straight away. It could be that many patients were being admitted who in retrospect only had mild acute pancreatitis. In that case, a bigger sample size would be needed to see the same effect on those patients with severe acute pancreatitis.

This means that the data monitoring committee was authorized to intervene in the design of the trial, for ethical reasons concerned with the treatment of future patients of future doctors, not with the treatment of their own patients in their own trial. They were authorized to make decisions about the future treatment of patients about to enter the trial, based on the results of the trial so far, without knowing which group was the treatment group and which group was the control group. It seems that the monitoring committee dealt with this problem by looking at the aggregate data of the two treatment groups. When they did this, they saw a much lower rate of infectious complications overall, than had been expected in advance, and hence concluded that there were more patients with only mild pancreatitis in the trial than planned. Simultaneously the monitoring committee saw the same overall mortality rate as had been expected in advance! It seems to me that an alarm bell might have gone off here. The monitoring committee does not know that the excess deaths -- not statistically significant, to be sure -- are occurring in the treatment group!  Despite the conflicting information, and blinded to the identity of the two groups, they proposed increasing the sample size.

Secondly, the application of Snapinn's rule talks about adverse events. Yet only two adverse events were ever investigated by the monitoring committee. Only two adverse events were ``serious''.  Both two serious adverse events were connected to the same complication. A literature search did not connect this kind of event to probiotica. What if the monitoring committee had known that already half of the many deaths in the probiotica group were of this same rare kind? Perhaps the literature search would have been extended with consultation with experts from other fields. It is now easy, after the events, to find microbiologists who say in effect ``I told you so''. 

I think it could have been good if the monitoring committee had talked to these microbiologists already half-way through the trial.

If the trial had not been triple-blinded, the monitoring committee might now have pulled the plug on it. Obviously (if the effect which has been found is real), this would have saved some lives of patients in the trial. 

Would science, would future patients of future doctors, have suffered? 

In retrospect there are plausible medical explanations for the ``new'' phenomenon. When your immune system is at breaking point, ``stimulating'' it with friendly bacteria is not a good idea. When the barriers between different organs are breaking down and evil, agressive, bacteria are moving freely from one place to another, adding new streams of migrants, however useful they might be in your gut, only makes things worse in places where they ought not to be.

It seems to this non-medical person, that if half-way the researchers had seen what was the special complication which was killing the already seriously ill patients in the treatment group, they could just as well have come up with this theory half-way already. It seems to this non-medical person, that if the trial had been stopped half-way, the substantive conclusions of the trial which now appear in \textit{The Lancet} could also have been reached and could also have been told to the world, one way or another.  Anyway, if the monitoring committee found other microbiologists with different ideas, they could have only temporarily stopped the trial. If it was medically truly believed to be a false alarm, the trial could have been recontinued after a break.

My point is, that at least they could, even if only in retrospect, have proved that they did always have the interests of their own patients in mind, not just the interests of the future patients of future doctors.
This would have made life for them easier now; and it would be to the benefit of future patients and of science, since clinical trials can only be done if the public is confident that their doctors always think in the first place of their own patients.

\section{The stopping rule}

As noted above, the PROPATRIA team report that the Snapinn stopping rule indicated that the trial should continue at the new (postponed) interim analysis, which actually took place not half-way through the modified trial at $150$, but a bit later at $184$ patients. It turns out that this conclusion was based on a mis-reading of the output of a statistical package, and that according to the protocol of the experiment the trial should have been stopped at that moment. In order to explain how this happened I need to discuss the stopping rule in a little more detail, and first of all to repeat the principles on which it is constructed.

The idea of the Snapinn rule is that a randomized clinical trial comparing an experimental new treatment to a standard therapy for a life-threatening medical condition should be stopped early on ethical grounds, in either of the following situations: (1) \emph{it has become overwhelmingly clear that the new treatment is better than the standard};
(2) \emph {it has become overwhelmingly clear that the trial is not going to show that the new treatment is any better than the standard}. These two situations are called ``stopping for significance'' and ``stopping for futility'', respectively.
The trial is continued in the third scenario: (3) \emph{there is a reasonable chance that the new treatment will finally turn out to be better than the standard, but we aren't sure yet}. 

An explicit possibility of stopping for futility provides an implicit safety measure: if the new treatment is actually harmful to the patients, we would want the trial to stop as early as possible. But this situation would also tend to result in data such that the trial is stopped early ``for futility''. 

The PROPATRIA team used an adaptation due to Schouten (1999, standard Dutch textbook \emph{Klinische Statistiek}) of the early-stopping-rule of Snapinn (1992, \emph{Statistics in Medicine}). The key feature of this early-stopping rule is that it is based on the $p$-value of the statistic of interest, at the time of the interim analysis. ``Time'' is expressed as the fraction of the originally planned sample size, at which one is performing the interim analysis. One should simply compare the interim $p$-value to two critical values, one for stopping for signficance, the other for stopping for futility. The critical values are determined from the over-all intended significance level and power, and the interim sample fraction (time). Snapinn's procedure is carefully designed such that the overall significance level of the trial with early stopping allowed is the same as the significance level of the trial with fixed sample size. Moreover, if the trial is not stopped early, the statistical analysis at the end of the trial is the same as if early stopping had not been incorporated in the design at all. Because the trial might stop early and the significance level is unaltered, some power is lost. Snapinn, and following him Schouten, have tuned the thresholds for early stopping in a compromise between loss of power and chance of stopping early. Schouten allows a bigger loss of power, hence increases the chance of stopping early.

As just mentioned, the data monitoring committee was blinded to the identity of treatment groups A and B. This means that they needed to compute the one-sided $p$-value for testing the null hypothesis of no treatment effect against the alternative of a beneficial treatment effect, with both assignments of ``treatment'' and ``control'' to groups A and B. The outcome is binary (infectious complications or not) and the researchers used the Fisher exact test for a $2\times 2$ contingency table. In that context, the statistical package SPSS does not allow the user to specify which one-sided alternative is of interest, but reports the $p$-value for the one-sided test which is more significant; i.e., that of the alternative suggested post-hoc by the data. Apparently, the committee did not realise what was going on. The $p$-value delivered by SPSS did not depend on the labelling of the two groups.

But what was it? Though the PROPATRIA researchers declined to provide the data from the interim analyses to interested scientists, they did accidentally provide some data to interested journalists at a press-conference. The two groups are labelled there as group A and group B. From the snippets of information about the interim analysis available in the Lancet paper, we can determine that group A is the treatment group, and group B is the placebo or control group. To my great surprise it turns out that at the interim analysis, the rate of infectious complications (the primary endpoint) in group A exceeded that in group B by an absolute amount of 5\%. Normalizing with an estimate of the of the standard deviation of the difference, yields a $z$-value of close to $+1$. Thus the one-sided $p$-value for the alternative that probiotica is \emph{bad} for you is about 16\%; the one-sided $p$-value for the alternative that probiotica is \emph{good} for you is about 84\%. 

The Fisher exact test gives similar $p$-values of 21\% and 87\% respectively. The data monitoring committee obtained from SPSS, for both cases, the smaller $p$-value of 21\%. Comparison with the critical values from Snapinn-Schouten leads to the advice ``continue'' in both cases. There is no need to de-blind the data. However the appropriate $p$-value was 87\% and the proper conclusion was to stop the experiment for futility.

In order to obtain the correct decision it would have been necessary to de-blind. The data monitoring committee would then not only have received a signal from the Snapinn rule that it was pointless to continue the trial, the rate of infectious complications in the treatment group was actually larger than in the control group and there was almost no chance that this could reverse by the end of the trial; they would also have seen that the mortality was also much larger in the treatment group than in the control group.

The fact that the researchers did not realise that anything was wrong indicates some lack of understanding of the principles behind the statistical methods they were using. There are some other indications of inadequate understanding, though to be fair, both Snapinn and Schouten-on-Snapinn are difficult reading. At the same time, it is very difficult to trace exactly what they did do: the publications of the probiotica group always refer to Snapinn (1992) without further specification. Snapinn gives three versions of this stopping rules, Schouten gives two more. The critical values quoted by the researchers cannot be found in Snapinn's paper at all. Now, Schouten made several innovations. He allows a greater chance of early stopping, at the cost of decrease of power. While Snapinn throughout works with a one-sided significance level of 2.5\%, Schouten uses throughout 5\%. There is a good reason for Snapinn's choice. He wants to graft his carefully constructed asymmetric early stopping rule onto a fixed sample size final evaluation with the conventional two-sided significance level of 5\%. Maybe in an instinctive correction for Schouten's inappropriate doubling of the significance level, the probiotica researchers took critical values corresponding to halving the error of the second kind: so together, they got their critical values from the table for one-sided 5\% significance level and power 90\%, instead of the ``trial design parameters''  one-sided 2.5\% signficance level and power 80\%.  There is one more mismatch: the interim analysis had been planned at a sample fraction of 50\%, but in fact only took place at 60\%; they should have gone back to the tables to find the critical values at 60\% rather than using those for 50\%.  However, none of these inaccuracies in the implementation affect the conclusion which should have followed: stop for futility.

\section{Optimal group sequential designs}

Jennison and Turnbull (2002) provide methodology for determining group sequential plans which minimize the expected sample size for given errors of the two kinds. In particular one can design a plan with the same errors of the first and second kinds as the PROPATRIA trial, and which minimizes the expected sample size when the actual effect of probiotica is the opposite to that expected Ð a doubling instead of a halving of the rate of infectious complications. It turns out that under that negative scenario, the expected sample size is 15\% of the fixed sample sample size, with the same size and power. 

However this result is not directly relevant to the PROPATRIA trial since the rate of infectious complications, the primary endpoint of the trial, was hardly affected by the treatment. Since infectious complications are the major cause of death in acute pancreatitis, and since deaths of some other causes also turned out to be increased by the treatment, in retrospect it would have been wise to take the primary endpoint of the trial as death from the disease. Now, the expected death rate was 10\%, and the researchers presumably expected this too to be halved by their treatment, while in fact it was doubled. Because of the lower rate, the fixed sample size required for two-sided size 5\% and power 80\% becomes larger. Still, the final result is that a group sequential plan designed for early stopping in the case of a negative effect of the treatment, would have led to this trial being stopped at about 100 patients.

\section{Conclusions}

Medical researchers following their hunches and anxious to prove they have found a fantastic new way to cure patients, pull sophisticated statistical tools out of the drawer. Using these sophisticated tools helps persuade ethical screening committees to back them. The standard methods in the standard textbooks already make ethical assumptions. Routine application of these methods means one is routinely making those same ethical assumptions. But I suspect that no-one realises what those ethical assumptions are. And no-one realises that addressing one ethical concern, might expose you more seriously to another. In principle mathematical statisticians can figure out the solutions to the more complicated optimization problems which arise when you try to take account of more, and conflicting, concerns at the same time. At least, we can bring these out into the open so everyone knows what is the cost to ``buying'' one of the standard solutions.

We could have designed the trial so that it minimized the expected number of patients entering the trial if the probiotica doubles the risk, subject to the same size and power. It would have been a completely different design. Possibly it would not have saved many lives, possibly it was infeasible. But I think it is important to know whether the researchers could easily have saved many lives, or if they were already doing close to what is best, even though they had not primarily concerned themselves with protecting their patients from a possibly dangerous medicine.

Admittedly a serious complication to any mathematical statistical treatment, is that in this case the treatment had increased the chance of a certain serious side effect. To be precise, death. So my main recommendation is not to do some difficult mathematics, but to reappraise the ethics of ``triple blind''. Especially on top of all the other actually rather one-sided ethical concern embodied in the usual choices of null and alternative, size and power.

The choice of the Snapinn protocol is one of the few places where the trial has a built in safety feature: if the treatment is working badly, there is some chance that ``stopping for futility'' will be triggered. It is especially tragic that this safety feature failed to work through a misreading of output of a statistical package.

My conclusion is that \textit{ethical screening} committees need more statistical expertise in order to judge which ethical concerns are being taken account of, when someone else's routine and technical ``ethical solution'' is implemented. Secondly, \textit{monitoring} committees also need more statistical expertise, in order to correctly implement complex statistical protocols. Thirdly, if the monitoring committee is blinded to the identity of the treatment and the control group, they should at least be advised by a person who is not blinded, in order to ensure that the committee never makes decisions based on an incorrect guess of which group is which.  Finally: it is important to learn from mistakes.

\section*{References}

\raggedright
\frenchspacing

M.G.H. Besselink, H.M. Timmerman, E. Buskens,  V.B. Nieuwenhuijs, L.M.A. Akkermans, 
H.G. Gooszen and the members of the Dutch Acute Pancreatitis Study Group (2004), 
Probiotic prophylaxis in patients with predicted severe acute 
pancreatitis (PROPATRIA): design and rationale of a double-blind, 
placebo-controlled randomised multicenter trial [ISRCTN38327949],
\textit{BMC Surgery}, 4:12 doi:10.1186/1471-2482-4-12 (7pp.)

\bigskip

M.G.H. Besselink et al. (2008), Probiotic prophylaxis in predicted severe acute pancreatitis: 
a randomised, double-blind, placebo-controlled trial,\\
\textit{The Lancet}, published online February 14, 2008, DOI:10.1016/S0140-6736(08)60207-X (9pp.)

\bigskip

H.J.A. Schouten (1999), \textit{Klinische Statistiek} (``Clinical Statistics''),\\
Houten: Bohn, Stafleu van Loghum.

\bigskip

S.M..Snapinn (1992), Monitoring clinical trials with a conditional probability stopping rule, 
\textit{Statistics in Medicine} \textbf{11}, 659--672.

\bigskip

J.P. Vandenbroucke (1999), Dwalingen in de methodologie XIV.\\
Het voortijdig be'indigen van een gerandomiseerde trial\\
(``Methodological errors XIV: Stopping a randomized trial too early''),\\
\textit{Nederlands Tijdschrift voor de Geneeskunde}  \textbf{143}, 1305--1308.

\end{document}